%% file: template-jems.tex
\documentclass{article}
\usepackage[journal=APDE,lang=british]{ems-journal-jems} 

\usepackage{textcomp}
\usepackage{mathrsfs}

\usepackage{chngcntr}
\usepackage{apptools}
\usepackage{verbatim}
\usepackage{tikz-cd}
\usepackage{listings}

\newtheorem{theorem}{Theorem}

\newtheorem{proposition}{Proposition}

\theoremstyle{definition}

\theoremstyle{remark}
\newtheorem{remark}{Remark}

\numberwithin{equation}{section}

\newcommand\restr[2]{{
		\left.\kern-\nulldelimiterspace 
		#1 
		\vphantom{\big|} 
		\right|_{#2} 
}}

\DeclareRobustCommand{\rchi}{{\mathpalette\irchi\relax}}
\newcommand{\irchi}[2]{\raisebox{\depth}{$#1\chi$}} 
\numberwithin{equation}{section}

\numberwithin{theorem}{section}
\numberwithin{proposition}{section}
\numberwithin{lemma}{section}
\numberwithin{remark}{section}
\numberwithin{exercise}{section}
\numberwithin{corollary}{section}

\newcounter{desccount}

\begin{document}

\title{Quadratic Bendixson criterion for delay systems: a numerical implementation}
\titlemark{Quadratic Bendixson criterion for delay systems: a numerical implementation}


\emsauthor{1}{
	\givenname{Mikhail}
	\surname{Anikushin}
	\orcid{0000-0002-7781-8551}}{M.M.~Anikushin}
\emsauthor{2}{
	\givenname{Andrey}
	\surname{Romanov}
	\orcid{0009-0008-4886-2351}}{A.O.~Romanov}

\Emsaffil{1}{
	\pretext{}
	\department{Department of Applied Cybernetics}
	\organisation{Faculty of Mathematics and Mechanics, St Petersburg University}
	\rorid{023znxa73}
	\address{Universitetskiy prospekt 28}
	\zip{198504}
	\city{Peterhof}
	\country{Russia}
	\posttext{}
	\affemail{demolishka@gmail.com}
}
\Emsaffil{2}{
	\pretext{}
	\department{Department of Applied Cybernetics}
	\organisation{Faculty of Mathematics and Mechanics, St Petersburg University}
	\rorid{023znxa73}
	\address{Universitetskiy prospekt 28}
	\zip{198504}
	\city{Peterhof}
	\country{Russia}
	\posttext{}
	\affemail{romanov.andrey.twai@gmail.com}
}
	

	


\classification[34K20, 34K27, 65-04]{37L15}
\keywords{delay equations; quadratic criterion; Bendixson criterion; global stability; integral operators}


\begin{abstract}
	We present a software implementation (as a Python package) of approximation schemes for verifying frequency inequalities arising from the quadratic Bendixson criterion for systems of delay equations (with a single delay in the linear part). In applications to dissipative nonlinear autonomous systems, such inequalities ensure global stability. The theoretical foundations of this approach were developed in our related work. In particular, it was shown that the transfer operators involved into the frequency inequalities can be represented as integral operators with bounded kernels of the Hardy--Volterra type. Here, we use this representation to construct approximation schemes based on finite-dimensional truncations of the transfer operators. Examples of using the package are provided.
\end{abstract}
\maketitle


\input{Introduction}
\input{Applications}
\input{Hopfield}





\begin{funding}
The reported study was funded by the Russian Science Foundation (Project 25-11-00147).
\end{funding}

\section*{Data availability}
The data that support the findings of this study can be generated using the scripts in the repository:\\
\centerline{\url{https://gitlab.com/romanov.andrey/quadratic-bendixson-dde}}

\section*{Conflict of interest}
The authors declare that they have no conflict of interest.


\end{document}

%% file: Introduction.tex
\section{Introduction}

In \cite{Smith1986HD}, Smith proposed a geometric approach to global stability problems for continuous-time systems with several equilibria based on a generalization of the Bendixson negative criterion in terms of attractors. This approach was further popularized by Li and Muldowney \cite{LiMuldowney1996SIAMGlobStab}, who also obtained a generalization of the criterion for systems in Banach spaces \cite{LiMuldowney1995LowBounds}. The classical line of reasoning in deducing global stability from the negative criterion relied on the upper semicontinuity of dimension estimates (see \cite{Anikushin2023LyapExp}), as well as on Pugh's closing lemma, whose infinite-dimensional variants are still unknown. In \cite{Anikushin2026Bendixson}, we succeeded in establishing global stability using purely topological arguments, thereby clarifying the real mechanisms underlying global stability and extending Smith's approach to infinite dimensions.

In addition to the cohomological constraints that always hold for global attractors in normed spaces, the quantitative part of the criterion assumes certain bounds for the dimension of the attractor. For smooth systems, the most effective approach to obtain such bounds is concerned with studying the evolution of infinitesimal volumes over the attractor. From a computational perspective (see \cite{Anikushin2023LyapExp}), the problem reduces to constructing a metric adapted to a specific class of problems, which allows for the detection of the exponential contraction of infinitesimal volumes. The largest dimension (possibly fractional) at which such contraction does not occur is called the \textit{Lyapunov dimension}. While for finite-dimensional systems there are results related to obtaining exact analytical formulas for the Lyapunov dimension (see \cite{Kuzetal2020Lorenz, Kuznetsov2016, LeoKuzKorKusakin2016, LeonovBoi1992, KuzReit2020}), in infinite dimensions most of the known explicit estimates are only asymptotically sharp (see \cite{Temam1997,ZelikAttractors2022,Anikushin2023LyapExp, AnikushinRomanov2024EffEst}).

In \cite{Anikushin2023Comp}, we proposed to study linear nonautonomous systems responsible for the evolution of infinitesimal volumes (compound cocycles) using the quadratic criterion from frequency-domain analysis (see \cite{Yakubovich1968,Yakubovich1998,Yakubovich2000}), and also laid functional-analytic foundations for applying this approach to delay equations\footnote{It appears that Mallet-Paret and Nussbaum were the first to explicitly investigate compound systems for certain scalar delay equations \cite{MalletParretNussbaum2013}.}, building on recent results concerning the frequency theorem \cite{Anikushin2020FreqDelay, Anikushin2020FreqParab, Likhtarnikov1977}. Geometrically, this approach, by means of a comparison with stationary linear systems, yields frequency-domain conditions in the form of inequalities that guarantee the existence of quadratic functionals defined on exterior powers, which are suitable for studying the contraction of infinitesimal volumes. However, the transfer operators appearing in such inequalities are purely infinite-dimensional, and their analysis cannot be carried out using rigorous mathematical methods alone.

In this regard, we developed numerical schemes in \cite{AnikushinRomanov2023FreqConds}. For scalar equations with a single delay in the distinguished linear block and their twofold extensions (responsible for the evolution of infinitesimal areas), it was observed that the transfer operators can be represented as integral operators with bounded kernels of the Hardy--Volterra type obtained through resolving a linear system of ODEs. In \cite{AnikushinRomanovQBC2026}, this approach was extended for systems of equations and termed the \textit{quadratic Bendixson criterion}. In this context, it is possible to derive highly efficient numerical schemes, which is the subject of the present work (see Section \ref{SEC: SystemsDelayCompoundOperators}).

Despite the great popularity of frequency-domain methods in the past, they have never been presented in a systematic way for investigating a wide range of problems in the context of infinite-dimensional systems. Our recent works aim to rectify this situation. This includes not only the studies of compound systems that have not been carried out before even in the case of ODEs, but also the development of a unified approach to inertial manifolds \cite{Anikushin2020Geom, Anikushin2022Semigroups, AnikushinAADyn2021, AnikushinRom2023SS}, which can be anticipated by the works of Smith \cite{Smith1992} and Miklav\v{c}i\v{c} \cite{Miclavcic1991}. Here, too, the quadratic criterion plays a central role.

The importance of frequency-domain methods for computational purposes is emphasized by the fact that they often produce better results than methods that formally appear more specialized (see \cite{SmirnovaProskurnikovZgoda2021, SmirnovaProskurnikovTitov2022, AnikushinRomanovQBC2026} and Section \ref{SEC: Examples}). This is because the mechanisms underlying applied models align with engineering intuition, which ensures that such models are not arbitrary dynamical systems. In particular, a significant gap between the region of global stability determined by a frequency criterion and the boundary of linear stability can often be considered as an indicator of the presence of nonlocal bifurcations in the system, which occur near or inside the linear stability boundary (see Section \ref{SEC: Examples}). The revealing of such scenarios is the topic of interest in the theory of hidden oscillations \cite{Kuznetsov2020,LeoKuz2013}, which has proven significant for many applied problems.

The structure of this paper is as follows. The main results are presented in Section \ref{SEC: SystemsDelayCompoundOperators}. First, we describe the class of equations to which the proposed framework is applicable and briefly outline the general approach to solving the problem (see Section \ref{SUBSEC: DelayClassStandingAssumptions}). Next, we examine the integral representation of transfer operators (see Section \ref{SUBSEC: IntegralRepresentationDelayCompound}). Based on this, we formulate frequency inequalities and investigate the regularity properties of the principal eigenfunction of the corresponding self-adjoint operator (see Section \ref{SUBSEC: FrequencyInequalitiesSubroutine}). Section \ref{SUBSEC: TruncationsOfTransferOperators} addresses finite-dimensional approximations (truncations) of integral operators, along with issues concerning their convergence and rigorous numerical analysis. The use of the software package is discussed in Section \ref{SUBSEC: PackageUtilization}. Finally, Section \ref{SEC: Examples} provides two computational examples that support the above discussion and demonstrate the consistency of the general computational scheme with our previous work.

%% file: Applications.tex
\section{Main results}
\label{SEC: SystemsDelayCompoundOperators}

\subsection{Description of the class of delay equations and the problem statement}
\label{SUBSEC: DelayClassStandingAssumptions}

We first describe the class of equations to which our methods are applicable.

Recall that a (continuous) \textit{semiflow} $\pi$ on a complete metric space $\mathcal{P}$ is a family of mappings $\pi^{t} \colon \mathcal{P} \to \mathcal{P}$, where $t \geq 0$, such that $\pi^{0}(\wp) = \wp$ and $\pi^{t+s}(\wp)=\pi^{t}(\pi^{s}(\wp))$ for any $\wp \in \mathcal{P}$, $t,s \geq 0$, and the mapping $[0,+\infty) \times \mathcal{P} \ni (t, \wp) \mapsto \pi^{t}(\wp) \in \mathcal{P}$ is continuous. Let $\mathbb{M} = \mathbb{R}^{d_{\mathbb{M}}}$ and $\mathbb{U} = \mathbb{R}^{d_{\mathbb{U}}}$ be Euclidean spaces\footnote{This means that the spaces $\mathbb{U}$ and $\mathbb{M}$ are equipped with inner products that are not necessarily the standard ones. The choice of inner products affects (quantitatively) the conditions concerning $F'(\wp)$ that will be presented below.} of dimensions $d_{\mathbb{M}}$ and $d_{\mathbb{U}}$, respectively. We study the following class of nonautonomous linear delay equations in $\mathbb{R}^{n}$ over $(\mathcal{P},\pi)$:
\begin{equation}
	\label{EQ: DelayNonautonomous}
	\dot{x}(t) = A_{0}x(t) + A_{\tau}x(t-\tau) + \widetilde{B}F'(\pi^{t}(\wp))C x_{t},
\end{equation}
where $\tau>0$ is the constant delay value, $x_{t}(\theta) \coloneq x(t+\theta)$ with $\theta \in [-\tau,0]$ is the $\tau$-history segment of $x(\cdot)$, $A_{0}$ and $A_{\tau}$ are $(n \times n)$-matrices, $C \colon C([-\tau,0];\mathbb{R}^{n}) \to \mathbb{M}$ is a bounded linear operator, $F'(\wp)$ is an $(d_{\mathbb{U}} \times d_{\mathbb{M}})$-matrix continuously depending on $\wp \in \mathcal{P}$, and $\widetilde{B}$ is an $(n \times d_{\mathbb{U}})$-matrix.

From the perspective of nonlinear systems, equations of the form \eqref{EQ: DelayNonautonomous} should be considered as the linearized equations, hence the name $F'$. Below, we will impose some further restrictions concerning $C$ and $F'$.

We assume that the measurement operator $C$ acts as follows:
\begin{equation}
	\label{EQ: MeasurementOperator}
	C\phi = \sum_{j=0}^{J+1} C_{\tau_{j}} \phi(-\tau_{j}) \qquad \text{for any} \ \phi \in C([-\tau,0];\mathbb{R}^{n}),
\end{equation}
where $C_{\tau_{1}}, \ldots, C_{\tau_{J+1}}$ are $(d_{\mathbb{M}} \times n)$-matrices and $0 = \tau_{0} < \tau_{1} < \cdots < \tau_{J+1} = \tau$.

There exists an integer $n_{\mathbb{U}} \geq 1$ and a splitting of $\mathbb{U}$ into the orthogonal sum,
\begin{equation}
	\mathbb{U} = \bigoplus_{i=1}^{n_{\mathbb{U}}} \mathbb{U}^{(i)},
\end{equation}
of certain subspaces $\mathbb{U}^{(i)}$. For each $\eta \in \mathbb{U}$, we write $\eta = (\eta_{1},\ldots,\eta_{n_{\mathbb{U}}})$, where $\eta_{i} \in \mathbb{U}^{(i)}$ for each $i$.

Next, for each $i \in \{1,\ldots, n_{\mathbb{U}}\}$, let there exist a subspace $\mathbb{M}^{(i)}$ of $\mathbb{M}$ such that for any $\wp \in \mathcal{P}$ we have that
\begin{equation}
	\label{EQ: DelayFeedbackComponentsDistinguished}
	F'(\wp) = (F'_{1}(\wp)\Pi^{(1)}_{\mathbb{M}}, \ldots, F'_{n_{\mathbb{U}}}(\wp)\Pi^{(n_{\mathbb{U}})}_{\mathbb{M}}),
\end{equation}
where $\Pi^{(i)}_{\mathbb{M}}$ is the orthogonal projector onto $\mathbb{M}^{(i)}$, and $F'_{i}(\wp)$ is a linear operator from $\mathbb{M}^{(i)}$ to $\mathbb{U}^{(i)}$. Note that the spaces $\mathbb{M}^{(i)}$ with different $i$ may intersect, or, in other words, different inputs may use the same measurements.

We assume that each $F'_{i}$ satisfies one of the following conditions\footnote{Possibly of different kind for distinct $i$.}:
\begin{description}[before=\let\makelabel\descriptionlabel]
	\item[\textbf{(LipC)}\refstepcounter{desccount}\label{DESC: LipC}] There exists a constant $\Lambda^{(i)} > 0$ such that $\|F'_{i}(\wp)\|_{\mathbb{M}^{(i)} \mapsto \mathbb{U}^{(i)}} \leq \Lambda^{(i)}$ for any $\wp \in \mathcal{P}$;
	\item[\textbf{(MonC)}\refstepcounter{desccount}\label{DESC: MonC}] The spaces $\mathbb{U}^{(i)}$ and $\mathbb{M}^{(i)}$ are isomorphic, and there exist linear mappings 
	\begin{equation}
		h_{i} \colon \mathbb{M} \to \mathbb{U} \quad \text{and} \quad h^{(-1)}_{i} \colon \mathbb{U} \to \mathbb{M}
	\end{equation}
	such that $h_{i}$ takes $\mathbb{M}^{(i)}$ isometrically onto $\mathbb{U}^{(i)}$, $h^{(-1)}_{i}$ takes $\mathbb{U}^{(i)}$ isometrically onto $\mathbb{M}^{(i)}$, and the restricted mappings are mutually inverse. Furthermore, there exist constants $\Lambda^{(i)}_{1} \leq 0 \leq \Lambda^{(i)}_{2}$ such that for any $\wp \in \mathcal{P}$, the operator $h^{(-1)}_{i} F'_{i}(\wp)$ is self-adjoint in $\mathbb{M}^{(i)}$ and satisfies the following inequality:
	\begin{equation}
		\Lambda^{(i)}_{1} \langle y, y \rangle_{\mathbb{M}} \leq \langle y, h^{(-1)}_{i} F'_{i}(\wp)y \rangle_{\mathbb{M}} \leq \Lambda^{(i)}_{2} \langle y, y \rangle_{\mathbb{M}} \quad \text{for any} \ y \in \mathbb{M}^{(i)}.
	\end{equation}
	In other words, the eigenvalues of $h^{(-1)}_{i} F'_{i}(\wp)$ lie in the interval $[\Lambda^{(i)}_{1},\Lambda^{(i)}_{2}]$. 
\end{description}
A concrete example that may help the reader understand these assumptions is provided in Section \ref{SUBSEC: HopfieldConcrete}.

Next, we will discuss frequency conditions related to the quadratic Bendixson criterion for \eqref{EQ: DelayNonautonomous}. Any such condition is a kind of balance inequality resulting in substituting the transfer operator of the linear block into the quadratic form defining the constraint associated with the feedback. Let us sketch a general perspective in our case. For this, it is required to rewrite \eqref{EQ: DelayNonautonomous} in an evolutionary form:
\begin{equation}
	\label{EQ: DelayLinearizedEvolutionary}
	\dot{\xi}(t) = A\xi(t) + BF'(\pi^{t}(\wp))C\xi(t),
\end{equation}
posed in an appropriate Hilbert space $\mathbb{H}$ and determined by the linear operators $A$, $B$, and $C$ associated with \eqref{EQ: DelayNonautonomous}. 

For the quadratic Bendixson criterion, we consider the extension of \eqref{EQ: DelayLinearizedEvolutionary} to the twofold exterior power $\mathbb{H}^{\wedge 2}$ of $\mathbb{H}$:
\begin{equation}
	\label{EQ: DelayLinearizedEvolutionaryTwofoldComp}
	\dot{\Phi}(t) = A^{[\wedge 2]} \Phi(t) + B^{\wedge}_{2}F^{\wedge}_{2}(\pi^{t}(\wp))C^{\wedge}_{2}\Phi(t),
\end{equation}
which is determined by the operators $A^{[\wedge 2]}$, $B^{\wedge}_{2}$, $C^{\wedge}_{2}$, and $F^{\wedge}_{2}$ associated with those in \eqref{EQ: DelayLinearizedEvolutionary}. So, $(A^{[\wedge 2]}, B^{\wedge}_{2}, C^{\wedge}_{2})$ and $F^{\wedge}_{2}$ determine the linear block and the feedback, respectively. We refer to \cite{AnikushinRomanovQBC2026} for precise definitions of the mentioned operators and involved spaces.

It turns out that the constraints imposed on $F'(\wp)$ naturally carry to $F^{\wedge}_{2}(\wp)$. This leads to the definition of a quadratic form $\mathcal{F}(M,\eta)$ of measurements $M$ and controls $\eta$ such that $\mathcal{F}(M,0) \geq 0$ and $\mathcal{F}(M,\eta) \geq 0$ provided that $\eta = F^{\wedge}_{2}(\wp)M$. In these terms, given $\nu_{0} \in \mathbb{R}$ such that the spectrum of $A^{[\wedge 2]}$ avoids the line $-\nu_{0} + i\mathbb{R}$, the frequency inequality requires for some $\delta > 0$ a with $\omega \in \mathbb{R}$ to satisfy the frequency inequality:
\begin{equation}
	\label{EQ: DelayTwoFoldFrequencyInequalityAbstract}
	\mathcal{F}^{\mathbb{C}}(W(-\nu_{0} + i \omega)\eta,\eta) \leq -\delta |\eta|^{2}_{\mathbb{U}^{\wedge}_{2}} \qquad \text{for all} \ \eta \in \mathbb{U}^{\wedge}_{2} \ \text{and} \ \omega \in \mathbb{R},
\end{equation}
where $\mathcal{F}^{\mathbb{C}}$ is the Hermitian extension of $\mathcal{F}$, $W(p) \coloneq C^{\wedge}_{2}(p I - A^{[\wedge 2]})^{-1}B^{\wedge}_{2}$ is the transfer operator associated with the triple $(A^{[\wedge 2]}, B^{\wedge}_{2}, C^{\wedge}_{2})$, and $|\cdot|_{\mathbb{U}^{\wedge}_{2}}$ is an appropriate norm in the (complexified) control space $\mathbb{U}^{\wedge}_{2}$. 

In this work, we develop numerical procedures for verifying \eqref{EQ: DelayTwoFoldFrequencyInequalityAbstract}. A more convenient form of this inequality is provided in Section \ref{SUBSEC: FrequencyInequalitiesSubroutine}, where relevant explanations are also presented.

\subsection{Integral representation of transfer operators}
\label{SUBSEC: IntegralRepresentationDelayCompound}
Next, we consider an integral representation of $W(p)$ obtained in \cite{AnikushinRomanovQBC2026}. For this, let $\mu$ be the measure on $[-\tau,0]$ given by the sum of the Lebesgue measure on $[-\tau,0]$ and the $\delta$-measure at $0$. If $\mathbb{F}$ is a Hilbert space, then any element $f$ of $L_{2}([-\tau,0];\mu;\mathbb{F})$ can be uniquely represented as a pair $(f_{1}(\cdot), f_{0})$, where $f_{1} \in L_{2}(-\tau,0;\mathbb{F})$ and $f_{0} \in \mathbb{F}$. In other words, this defines a natural isometric isomorphism between $L_{2}([-\tau,0];\mu;\mathbb{F})$ and the product space $L_{2}(-\tau,0;\mathbb{F}) \times \mathbb{F}$. This will be used in what follows.

Let $I_{n}$ be the identity operator in $\mathbb{C}^{n}$. Then $I^{\otimes 2}_{n} = I_{n} \otimes I_{n}$ is the identity operator in $(\mathbb{C}^{n})^{\otimes 2} = \mathbb{C}^{n} \otimes \mathbb{C}^{n}$. Set also $B^{1}_{2} = B^{0}_{2} = I_{n} \otimes \widetilde{B}$, and let $T_{(12)}$ be the transposition operator in $(\mathbb{C}^{n})^{\otimes 2}$:
\begin{equation}
	T_{(12)}(x \otimes y) = y \otimes x \quad \text{for any} \ x,y \in \mathbb{C}^{n}.
\end{equation}
If $L$ is an operator, we set $L^{2} \coloneq I_{n} \otimes L$ and $L^{1} \coloneq L \otimes I_{n}$. This will be used for $L \in \{ A_{0}, A_{\tau}, C_{\tau_{0}}, \ldots, C_{\tau_{J+1}} \}$, thereby defining operators $A^{1}_{0}$, $A^{2}_{0}$, $A^{1}_{\tau}$, $A^{2}_{\tau}$, and $C^{2}_{\tau_{j}}$ for $j \in \{0,\ldots, J+1\}$.

According to \cite[Section II]{AnikushinRomanovQBC2026}, the action of the transfer operator $W(p)$ reduces to the following controls-to-measurements correspondence:
\begin{equation}
	\label{EQ: TwofoldDelayCompoundWCorrespondence}
	\mathbb{H}_{\mathbb{U}} \ni (\eta^{2}_{1}, \eta^{2}_{0}) \mapsto (M^{2}_{1}, M^{2}_{0}) \in \mathbb{H}_{\mathbb{M}},
\end{equation}
where $\mathbb{H}_{\mathbb{U}} \coloneq L_{2}([-\tau,0]; \mu; \mathbb{C}^{n} \otimes \mathbb{U})$, $\mathbb{H}_{\mathbb{M}} \coloneq L_{2}([-\tau,0]; \mu; \mathbb{C}^{n} \otimes \mathbb{M})$,
\begin{equation}
	\label{EQ: TwofoldDelayIntegralMeasurementClosure}
	\begin{split}
		M^{2}_{1}(\theta) &= \sum_{j=0}^{J+1}\begin{cases}
			-e^{-\tau_{j} p}C^{2}_{\tau_{j}} y_{1}(\theta + \tau_{j}) \quad &\text{if} \ \theta \in [-\tau,-\tau_{j}], \\
			e^{p\theta} C^{2}_{\tau_{j}}T_{(12)}y_{1}(-\tau_{j}-\theta) \quad &\text{if} \ \theta \in [-\tau_{j}, 0].
		\end{cases},\\
		M^{2}_{0} &= \sum_{j=0}^{J+1}C^{2}_{\tau_{j}}T_{(12)}y_{1}(-\tau_{j}),
	\end{split}
\end{equation}
where $y_{1}(\cdot) \in W^{1,2}(-\tau,0;(\mathbb{C}^{n})^{\otimes 2})$ is a solution on $[-\tau,0]$ to the following linear system of ODEs in $(\mathbb{C}^{n})^{\otimes 2} \times (\mathbb{C}^{n})^{\otimes 2}$:
\begin{equation}
	\label{EQ: ExplicitTwofoldResolventEquationSystemY1Y2DiffEq}
	\begin{split}
		y'_{1} &= (pI^{\otimes 2}_{n} - A^{2}_{0}) y_{1} + e^{p\theta} A^{2}_{\tau} T_{(12)} y_{2} + \psi(\theta),\\
		y'_{2} &= (A^{2}_{0} - pI^{\otimes 2}_{n}) y_{2} - e^{-p(\tau+\theta)} A^{2}_{\tau} T_{(12)} y_{1} - \psi(-\tau-\theta),
	\end{split}
\end{equation}
satisfying the boundary conditions\footnote{The boundary condition $y_{1}(-\tau) = y_{2}(0)$ follows from the previous ones.}:
\begin{equation}
	\label{EQ: ExplicitTwofoldResolventEquationSystemY1Y2Boundary1}
	\begin{split}
		\left(A^{1}_{0} + A^{2}_{0} - p I^{\otimes 2}_{n}\right) y_{1}(0) + \left( I^{\otimes 2}_{n} - T_{(12)} \right)A^{1}_{\tau}y_{2}(0) = \psi(0),\\
		y_{1}(0)=y_{2}(-\tau) \quad \text{and} \quad y_{1}(-\tau)=y_{2}(0),
	\end{split}
\end{equation}
and $\psi(\cdot)$ and $\psi(0)$ are given by
\begin{equation}
	\label{EQ: PsiResolventControlClosure}
	\psi(\cdot) = B^{1}_{2} \eta^{2}_{1} \quad \text{and} \quad \psi(0) = (I^{\otimes 2}_{n} - T_{(12)})B^{0}_{2}\eta^{2}_{0}.
\end{equation}
\begin{remark}
	For convenience, we do not distinguish between $\mathbb{U}$ and $\mathbb{M}$ and their complexifications used in the tensor products $\mathbb{C}^{n} \otimes \mathbb{U}$ and $\mathbb{C}^{n} \otimes \mathbb{M}$, respectively. Note also that $\psi(\cdot)$ and $\psi(0)$ are unrelated in general.
\end{remark}

Next, we turn to an explicit integral representation of the correspondence from \eqref{EQ: TwofoldDelayCompoundWCorrespondence}. According \cite[Lemma II.1]{AnikushinRomanovQBC2026}, there exists a unique solution $(y_{1}, y_{2})$ to \eqref{EQ: ExplicitTwofoldResolventEquationSystemY1Y2DiffEq}-\eqref{EQ: ExplicitTwofoldResolventEquationSystemY1Y2Boundary1}. It also satisfies the symmetry $y_{2}(\theta) = y_{1}(-\tau-\theta)$ for any $\theta \in [-\tau,0]$. Moreover, according to \cite[Lemma II.2]{AnikushinRomanovQBC2026}, the operators
\begin{equation}
	\label{EQ: SolvingCauchyOperators}
	\begin{split}
	G_{D}(\theta) &\coloneq \begin{pmatrix}
		e^{p\theta} I^{\otimes 2}_{n} & 0\\
		0 &  I^{\otimes 2}_{n}
	\end{pmatrix}
	e^{D_{0}(\theta+\tau)}
	\begin{pmatrix}
		e^{p\tau} I^{\otimes 2}_{n} & 0\\
		0 & I^{\otimes 2}_{n}
	\end{pmatrix}, \quad \text{where}\\
	 D_{0} &\coloneq \begin{pmatrix}
		-A^{2}_{0} & A^{2}_{\tau}T_{(12)}\\
		-e^{-p\tau}A^{2}_{\tau}T_{(12)} & A^{2}_{0}-pI^{\otimes 2}_{n}
	\end{pmatrix},
	\end{split}
\end{equation}
are the Cauchy resolving operators for the homogeneous problem, that is, \eqref{EQ: ExplicitTwofoldResolventEquationSystemY1Y2DiffEq} with $\psi \equiv 0$. Consequently, the general solution to
\eqref{EQ: ExplicitTwofoldResolventEquationSystemY1Y2DiffEq} has the following form:
\begin{equation}
	\label{EQ: ExResolventEqsGeneralInhFormula}
	\begin{pmatrix}
		y_{1}(\theta)\\
		y_{2}(\theta)
	\end{pmatrix}
	= G_{D}(\theta) 		\begin{pmatrix}
		y_{1}(-\tau)\\
		y_{2}(-\tau)
	\end{pmatrix}
	+ \int_{-\tau}^{\theta} G_{D}(\theta) (G_{D}(s))^{-1} \begin{pmatrix}
		\psi(s)\\
		-\psi(-\tau-s)
	\end{pmatrix}
	ds.
\end{equation}

To find the solution satisfying the boundary conditions, it is convenient to assume that $p$ does not belong to the spectrum of the operator $A^{[\otimes 2]} = A^{1}_{0} + A^{2}_{0}$, that is, $p$ cannot be represented as the sum of two (possibly repeated) eigenvalues of $A_{0}$. Thus, $J_{0} \coloneq (A^{1}_{0} + A^{2}_{0} - p I^{\otimes 2}_{n})^{-1}$ is well-defined. From this, we can rewrite the first equation in \eqref{EQ: ExplicitTwofoldResolventEquationSystemY1Y2Boundary1} as follows: 
\begin{equation}
	\label{EQ: ResolventY10}
	y_{1}(0) = J_{1} y_{1}(-\tau) + J_{0}\psi(0),
\end{equation}
where we used that $y_{2}(0) = y_{1}(-\tau)$ and set $J_{1} \coloneq  J_{0} (T_{(12)} - I^{\otimes 2}_{n})A^{1}_{\tau}$. 

Writing $G_{D}(\theta)$ and $G_{D}(\theta)(G_{D}(s))^{-1}$ as operator matrices:
\begin{equation*}
	\label{EQ: TwofoldDelayOperatorMatricesComponentRepr}
	G_{D}(\theta) = \begin{pmatrix}
		G_{11}(\theta) & G_{12}(\theta)\\
		G_{21}(\theta) & G_{22}(\theta)
	\end{pmatrix}
	\quad \text{and} \quad
	G_{D}(\theta)(G_{D}(s))^{-1} = \begin{pmatrix}
		G_{11}(\theta, s) & G_{12}(\theta, s)\\
		G_{21}(\theta, s) & G_{22}(\theta, s)
	\end{pmatrix},
\end{equation*}
we obtain the following equations for $y_{1}(-\tau)$ by substituting $\theta = 0$ into \eqref{EQ: ExResolventEqsGeneralInhFormula} and using \eqref{EQ: ResolventY10} and \eqref{EQ: ExplicitTwofoldResolventEquationSystemY1Y2Boundary1}:
\begin{equation}
	\label{EQ: ResolventTwofoldDelayY1tau}
	\begin{split}
		\begin{pmatrix}
			J_{1}y_{1}(-\tau) + J_{0}\psi(0)\\
			y_{1}(-\tau)
		\end{pmatrix}
		=
		\begin{pmatrix}
			G_{11}(0) y_{1}(-\tau) + G_{12}(0)(J_{1}y_{1}(-\tau) + J_{0}\psi(0))\\
			G_{21}(0) y_{1}(-\tau) + G_{22}(0) (J_{1}y_{1}(-\tau) + J_{0}\psi(0))
		\end{pmatrix}\\
		+
		\int_{-\tau}^{0} \begin{pmatrix}
			(G_{11}(0,s)- G_{12}(0,-\tau-s))\psi(s)\\
			(G_{21}(0,s) - G_{22}(0,-\tau-s))\psi(s)
		\end{pmatrix}ds,
	\end{split}
\end{equation}
where we also applied a change of variables:
\begin{equation}
	\label{EQ: G1222ChangeOfvar}
	\int_{-\tau}^{0} \begin{pmatrix}
		G_{12}(0,s)\psi(-\tau-s)\\
		G_{22}(0,s)\psi(-\tau-s)
	\end{pmatrix}ds = \int_{-\tau}^{0} \begin{pmatrix}
		G_{12}(0,-\tau-s)\psi(s)\\
		G_{22}(0,-\tau-s)\psi(s)
	\end{pmatrix}ds.
\end{equation}

By our choice of $p$, we are guaranteed to have a unique solution $y_{1}(-\tau)$ of \eqref{EQ: ResolventTwofoldDelayY1tau} for any $\psi(\cdot)$ and $\psi(0)$. We do not know whether any difficulties might arise in finding $y_{1}(-\tau)$ just by inversion from any of the componentwise equations. Theoretically, such a possibility would lead to strong degeneracy of components of the operator matrix exponential. Since these are analytic functions of $p$, this should occur extremely rarely. We have also never encountered such problems in numerical experiments. So, we assume that the operator
\begin{equation}
	G_{y_{1}} = (I^{\otimes 2}_{n} - G_{12}(0))J_{1} - G_{11}(0)
\end{equation}
is invertible. In any case, the invertibility of $G_{y_{1}}$ can easily be verified experimentally and made rigorous with the use of interval arithmetic.

From this, equating the first components in \eqref{EQ: ResolventTwofoldDelayY1tau}, we can express $y_{1}(-\tau)$  as follows:
\begin{equation}
	\label{EQ: Y1tauIntRep}
	\begin{split}
	y_{1}(-\tau) &= \int_{-\tau}^{0}Y_{01}(s)\psi(s)ds + Y_{00} \psi(0), \quad \text{where}\\
	Y_{01}(s) &= G_{y_{1}}^{-1} ( G_{11}(0,s) - G_{12}(0,-\tau-s) ),\\
	Y_{00} &= G_{y_{1}}^{-1} ( G_{12}(0) - I^{\otimes 2}_{n} )J_{0}.
	\end{split}
\end{equation}

Next, from \eqref{EQ: ResolventY10} we can recover $y_{2}(-\tau) = y_{1}(0)$ and similarly obtain that
\begin{equation}
	\label{EQ: Y2tauIntRep}
	\begin{split}
		y_{2}(-\tau) &= \int_{-\tau}^{0}Y'_{01}(s)\psi(s)ds + Y'_{00} \psi(0), \quad \text{where}\\
		Y'_{01}(s) &= J_{1}Y_{01}(s) \quad \text{and} \quad Y'_{00} = J_{1}Y_{00} + J_{0}.
	\end{split}
\end{equation}

Substituting \eqref{EQ: Y1tauIntRep} and \eqref{EQ: Y2tauIntRep} into \eqref{EQ: ExResolventEqsGeneralInhFormula} gives the following representation of $y_{1}$:
\begin{equation}
	\label{EQ: Y1TwofoldResolventIntegralFormula}
	\begin{split}
		y_{1}(\theta) &= \int_{-\tau}^{0}Y_{11}(\theta, s) \psi(s)ds + Y_{10}(\theta) \psi(0), \quad \text{where}\\
		Y_{11}(\theta, s) &= G_{11}(\theta)Y_{01}(s) + G_{12}(\theta)Y'_{01}(s) \\&+ \rchi_{[-\tau,\theta]}(s)G_{11}(\theta,s) - \rchi_{[-\tau-\theta,0]}(s)G_{12}(\theta,-\tau-s),\\
		Y_{10}(\theta) &= G_{11}(\theta)Y_{00} + G_{12}(\theta)Y'_{00},
	\end{split}
\end{equation}
where we also applied a change of variables similarly to \eqref{EQ: G1222ChangeOfvar}, which resulted in the characteristic function $\rchi_{[-\tau-\theta,0]}(s)$. Note that the inner kernel $Y_{11}(\theta,s)$ generally has discontinuities along the lines $\theta=s$ and $\theta = -\tau - s$ due to the presence of characteristic functions. If $A_{\tau}$ is zero, then $G_{D}$ from \eqref{EQ: SolvingCauchyOperators} is diagonal, and all the terms with $G_{12}$ vanish.

Closing $(\psi(\cdot),\psi(0))$ via controls $(\eta^{2}_{1}, \eta^{2}_{0})$ as in \eqref{EQ: PsiResolventControlClosure} and then constructing the pair of measurements $(M^{2}_{1}, M^{2}_{0})$ from $y_{1}$ via \eqref{EQ: TwofoldDelayIntegralMeasurementClosure}, we obtain that the correspondence \eqref{EQ: TwofoldDelayCompoundWCorrespondence} is delivered by an integral operator $T_{K}$ with a bounded operator kernel $K$ of the Hardy--Volterra type\footnote{That is, the kernel possesses discontinuities resulting from the presence of characteristic functions in the corresponding formulas.}. Similarly to the above, it is convenient to represent $K$ through its 4 components: $K(\cdot_{1},\cdot_{2})$, $K(\cdot,0)$, $K(0,\cdot)$, and $K(0,0)$. In these terms, we have the following:
\begin{equation}
	\label{EQ: TransferOperatorTwoFoldDelayIntegralRepresentation}
	\begin{split}
		M^{2}_{1}(\theta) = \int_{-\tau}^{0}K(\theta,s)\eta^{2}_{1}(s)ds + K(\theta,0)\eta^{2}_{0},\\
		M^{2}_{0} = \int_{-\tau}^{0}K(0,s)\eta^{2}_{1}(s)ds + K(0,0) \eta^{2}_{0}.
	\end{split}
\end{equation}

Note that each measurement from \eqref{EQ: TwofoldDelayIntegralMeasurementClosure} contributes to the discontinuities of $K$ in the following way. While boundary delays $\tau_{j} \in \{0, \tau\}$ preserve the discontinuity lines of $Y_{11}(\theta,s)$ for $K(\theta,s)$, the inner delays $\tau_{j} \in (0,\tau)$  shift them, thereby creating new discontinuities. Moreover, the inner delays also force the boundary kernel $K(0,s)$ to have discontinuities. This should be taken into account when performing numerical integration.
\begin{remark}
	When $n=1$, we do not need to consider the $0$-boundary components $\eta^{2}_{0}$ and $M^{2}_{0}$. This is because they affect only the $0$-boundary component of \eqref{EQ: DelayLinearizedEvolutionaryTwofoldComp}, which belongs to $(\mathbb{R}^{n})^{\wedge 2}$ and hence is identically zero for $n = 1$ (see \cite{AnikushinRomanovQBC2026,Anikushin2023Comp,AnikushinRomanov2023FreqConds} for details). From the general perspective, we have that $\psi(0)$ in \eqref{EQ: PsiResolventControlClosure} closes by $0$ in this case, since $I^{\otimes 2}_{n} - T_{(12)} = 0$. Consequently, the components $K(\theta,0)$ and $K(0,0)$ vanish. Thus, only $K(0,s)$ should be forced to zero.
\end{remark}

\subsection{Frequency inequalities associated with the feedback constraints}
\label{SUBSEC: FrequencyInequalitiesSubroutine}
Next, we discuss how the frequency inequalities from \eqref{EQ: DelayTwoFoldFrequencyInequalityAbstract} are expressed in terms of $T_{K}=T_{K}(p)$.
\begin{remark}
	\label{REM: UnitaryEquivalenceWandTK}
	For details we refer to \cite[Section 6.4]{Anikushin2023Comp}, where the frequency inequalities are derived in terms of $W(p)$. In our case, $W(p)$ takes control quadruples $(\eta^{2}_{1}, \eta^{1}_{2}, \eta^{1}_{0}, \eta^{2}_{0}) \in \mathbb{U}^{\wedge}_{2}$ into measurement quadruples $(M^{2}_{1}, M^{1}_{2}, M^{1}_{0}, M^{2}_{0}) \in \mathbb{M}^{\wedge}_{2}$, which satisfy the antisymmetric relations:
	\begin{equation}
		\eta^{1}_{2} = -T_{(12)} \eta^{2}_{1}, \eta^{1}_{0} = -T_{(12)}\eta^{2}_{0}, M^{1}_{2} = -T_{(12)} M^{2}_{1}, \ \text{and} \ M^{1}_{0} = -T_{(12)}M^{2}_{0},
	\end{equation}
	where $T_{(12)}$ transposes $\mathbb{C}^{n} \otimes \mathbb{U}$ (resp. $\mathbb{C}^{n} \otimes \mathbb{M}$) to $\mathbb{U} \otimes \mathbb{C}^{n}$ (resp. $\mathbb{M} \otimes \mathbb{C}^{n}$). By virtue of this antisymmetry, $W(p)$ is unitarily equivalent to $T_{K}(p)$, which is just the correspondence $(\eta^{2}_{1},\eta^{2}_{0}) \mapsto (M^{2}_{1}, M^{2}_{0})$.
\end{remark}

Recall that we distinguished the components $F'_{i}(\wp) \colon \mathbb{M}^{(i)} \to \mathbb{U}^{(i)}$ of the feedback $F'(\wp)$ in \eqref{EQ: DelayFeedbackComponentsDistinguished}, and each component $F'_{i}(\wp)$ satisfies one of the constraints \nameref{DESC: LipC} or \nameref{DESC: MonC}. From this, we naturally associate with $\mathbb{U}^{(i)}$ and $\mathbb{M}^{(i)}$ the subspaces $\mathbb{U}^{(i)}_{2}$ and $\mathbb{M}^{(i)}_{2}$ in $\mathbb{H}_{\mathbb{U}}$ and $\mathbb{H}_{\mathbb{M}}$, respectively. Specifically, $(\eta^{2}_{1}, \eta^{2}_{0}) \in \mathbb{U}^{(i)}_{2}$ if and only if $\eta^{2}_{1}(\theta), \eta^{2}_{0} \in \mathbb{C}^{n} \otimes \mathbb{U}^{(i)}$ for almost all $\theta \in (-\tau,0)$, and similarly for $\mathbb{M}^{(i)}_{2}$.

Let $\Pi^{(i)}_{\mathbb{U}^{\wedge}_{2}}$ and $\Pi^{(i)}_{\mathbb{M}^{\wedge}_{2}}$ be the orthogonal projectors onto $\mathbb{U}^{(i)}_{2}$ and $\mathbb{M}^{(i)}_{2}$, respectively. Similarly, let $F^{\wedge}_{m,i}(\wp)$ denote the mapping from $\mathbb{M}^{(i)}_{m}$ to $\mathbb{U}^{(i)}_{m}$  induced by $F'_{i}(\wp)$, and let $h_{m, i}$ (resp. $h^{(-1)}_{m, i}$) denote the induced by $h_{i}$ (resp.  $h^{(-1)}_{i}$) from \nameref{DESC: MonC} linear mapping that takes $\mathbb{M}^{(i)}$ (resp. $\mathbb{U}^{(i)}$) into $\mathbb{U}^{(i)}$ (resp. $\mathbb{M}^{(i)}$).

From this, we define a self-adjoint operator $S_{i}(p)$ in $\mathbb{H}_{\mathbb{U}}$ either given by
\begin{equation}
	\label{EQ: DelayCompoundSiLipschitz}
	S_{i}(p) = |\Lambda^{(i)}|^{2}T^{*}_{K}(p) \Pi^{(i)}_{\mathbb{M}^{\wedge}_{2}} T_{K}(p) - \Pi^{(i)}_{\mathbb{U}^{\wedge}_{2}},
\end{equation}
if $F'_{i}$ satisfies \nameref{DESC: LipC}, or by
\begin{equation}
	\label{EQ: DelayCompoundSiMonotone}
	\begin{split}
		S_{i}(p) = -\Lambda^{(i)}_{1} \Lambda^{(i)}_{2} T^{*}_{K}(p) \Pi^{(i)}_{\mathbb{M}^{\wedge}_{2}} T_{K}(p) +\\
		\frac{1}{2}(\Lambda^{(i)}_{1} + \Lambda^{(i)}_{2})\left(T^{*}_{K}(p) \Pi^{(i)}_{\mathbb{M}^{\wedge}_{2}} h^{(-1)}_{2,i} \Pi^{(i)}_{\mathbb{U}^{\wedge}_{2}} + \Pi^{(i)}_{\mathbb{U}^{\wedge}_{m}} h_{2, i} \Pi^{(i)}_{\mathbb{M}^{\wedge}_{2}} T_{K}(p)\right) - \Pi^{(i)}_{\mathbb{U}^{\wedge}_{2}},
	\end{split}
\end{equation}
if $F'_{i}$ satisfies \nameref{DESC: MonC}. 

Given $\kappa = (\kappa_{1},\ldots,\kappa_{d_{\mathbb{U}}})$ with $\varkappa_{i} \geq 0$ for $i \in \{1,\ldots,n_{\mathbb{U}}\}$, we set
\begin{equation}
	\label{EQ: DelayCompoundSigmaKappaSum}
	\Sigma^{(\kappa)}(p) \coloneq \sum_{i=1}^{n_{\mathbb{U}}}\kappa_{i} S_{i}(p).
\end{equation}
Then the frequency inequality takes the following form:
\begin{equation}
	\label{EQ: FrequencyInequalityCompoundDelayMixedConstraint}
	\sup_{\omega \in \mathbb{R}}\lambda_{\operatorname{max}}( \Sigma^{(\kappa)}(-\nu_{0} + i \omega) ) < 0,
\end{equation}
where $\lambda_{\operatorname{max}}$ is the largest eigenvalue, which is the upper spectral bound of the operator enclosed in the parentheses. According to Remark \ref{REM: UnitaryEquivalenceWandTK}, the condition \eqref{EQ: FrequencyInequalityCompoundDelayMixedConstraint} is precisely \cite[Equation (6.76)]{Anikushin2023Comp}. In practice, the real number $\nu_{0}$ in \eqref{EQ: FrequencyInequalityCompoundDelayMixedConstraint} is chosen as follows.

The spectrum of the operator $A$ in \eqref{EQ: DelayLinearizedEvolutionary}, which is associated with the matrices $A_{0}$ and $A_{\tau}$ in \eqref{EQ: DelayNonautonomous}, consists of eigenvalues determined from the characteristic equation:
\begin{equation}
	\label{EQ: DelayOperatorCharacEq}
	\operatorname{det}\left( A_{0} + e^{-p \tau}A_{\tau} - pI_{n} \right) = 0,
\end{equation}
where, as above, $I_{n}$ denotes the identity $n \times n$-matrix. Enumerating the roots of \eqref{EQ: DelayOperatorCharacEq} by decreasing real parts and according to their multiplicity, let $\lambda_{1}(A)$ and $\lambda_{2}(A)$ be the leading pair of them. Then $s(A^{[\wedge 2]}) = \lambda_{1}(A) + \lambda_{2}(A)$ determines the spectral bound of the operator $A^{[\wedge 2]}$ in \eqref{EQ: DelayLinearizedEvolutionaryTwofoldComp}.

\begin{theorem}
	\label{TH: ExpontialStabilityTwofoldDelayGeneral}
	Let the assumptions of Section \ref{SUBSEC: DelayClassStandingAssumptions} hold for \eqref{EQ: DelayNonautonomous}. Suppose that $s(A^{[\wedge 2]}) < 0$ and \eqref{EQ: FrequencyInequalityCompoundDelayMixedConstraint} is valid for some $\nu_{0} \in [0,-s(A^{[\wedge 2]}))$ and $\varkappa_{i} \geq 0$, where $i \in \{1,\ldots,n_{\mathbb{U}}\}$. Then \eqref{EQ: DelayLinearizedEvolutionaryTwofoldComp} is uniformly exponentially stable with the exponent $\nu_{0} + \varepsilon$ for all sufficiently small $\varepsilon > 0$.
\end{theorem}
\begin{proof}
	This is a special case of \cite[Theorem 6.2]{Anikushin2023Comp} subject to the same remarks as those provided in \cite[Theorem II.1]{AnikushinRomanovQBC2026}.
\end{proof}
\begin{remark}
	Recall that if \eqref{EQ: DelayNonautonomous} is obtained by linearizing a nonlinear autonomous dissipative system, then the uniform exponential stability of \eqref{EQ: DelayLinearizedEvolutionaryTwofoldComp} guarantees the global stability of \eqref{EQ: DelayNonautonomous}, see \cite[Theorem IV.1]{AnikushinRomanovQBC2026}. For certain nonautonomous systems, there also exist constraints related to area contraction, see \cite{Anikushin2026Bendixson}.
\end{remark}

For a general relationship between the spectra of $A$ and $A^{[\wedge 2]}$, see \cite[Theorems 3.2 and 3.3]{Anikushin2023Comp}. If $-\nu_{0}$ is chosen such that the line $-\nu_{0} + i\mathbb{R}$ separates the first $j$ eigenvalues of $A^{[\wedge 2]}$ from the rest, then the validity of \eqref{EQ: FrequencyInequalityCompoundDelayMixedConstraint} guarantees the existence of a uniform exponential semi-dichotomy (or dichotomy, if $(\mathcal{P},\pi)$ is invertible) for \eqref{EQ: DelayLinearizedEvolutionaryTwofoldComp}, which is a perturbation of the corresponding spectral subspaces of $A^{[\wedge 2]}$.

From \eqref{EQ: FrequencyInequalityCompoundDelayMixedConstraint}, it is clear that without loss of generality we can assume that all $\kappa_{i}$ form a convex combination. Of course, the ideal choice of such parameters is to minimize the left-hand side of \eqref{EQ: FrequencyInequalityCompoundDelayMixedConstraint}. This can be achieved numerically with sufficient accuracy using truncations of $T_{K}$, which we discuss in the next section. In theses terms, the optimization problem falls into the realm of nonlinear programming, where it can be solved efficiently.
\begin{remark}
	Since we deal with the complexification of a real problem, the quantities $\alpha(\omega)$ under the supremum in \eqref{EQ: FrequencyInequalityCompoundDelayMixedConstraint} are symmetric in $\omega$: $\alpha(\omega) = \alpha(-\omega)$. Thus, the supremum can be taken only over $\omega \geq 0$.
\end{remark}
\begin{remark}
	\label{REM: DelayScalingNoninvariance}
	For systems of equations, frequency inequalities are not invariant under the delay scaling $t \mapsto \varkappa t$ in \eqref{EQ: DelayNonautonomous} with $\varkappa > 0$, see \cite{AnikushinRomanovQBC2026}. In a number of situations, choosing an appropriate scaling can significantly improve the results obtained.
\end{remark}

At the conclusion of this subsection, we establish the smoothness properties of the principal eigenfunction of the operator $\Sigma^{(\kappa)}(p)$. This will be used to determine the rate of convergence of approximations to \eqref{EQ: FrequencyInequalityCompoundDelayMixedConstraint} delivered by truncations in appropriate bases.
\begin{proposition}
	\label{PROP: PrincipalEigenfunctionDelayCompound}
	Let $f=(f_{1},f_{0}) \in \mathbb{H}_{\mathbb{U}}$ be the eigenfunction of $\Sigma^{(\kappa)}(p)$ corresponding to the largest eigenvalue $\lambda$, and suppose that $\lambda + \kappa_{i} \not= 0$ for any $i \in \{1, \ldots, d_{\mathbb{U}}\}$. Then $f_{1}$ belongs to $C([-\tau,0];\mathbb{C}^{n} \otimes \mathbb{U})$ and is piecewise $C^{1}$-smooth with possible jumps of the derivative at the points $-\tau - \tau_{j}$ and $-\tau_{j}$ corresponding to the inner delays $-\tau_{j} \in (-\tau,0)$. Moreover, $f_{1}$ is piecewise $C^{k}$-smooth, with the set of points of discontinuity for its derivatives (up to order $k$) being finite and constructed inductively with respect to $k$ in an appropriate manner, depending on the inner delays. In particular, in the absence of inner delays, $f_{1}$ is $C^{\infty}$-smooth.
\end{proposition}
\begin{proof}
	First, we represent $T_{K}$ as the following composition of operators:
	\begin{equation}
		(\eta^{2}_{1},\eta^{2}_{0}) \overset{T_{B}}{\mapsto} (\psi(\cdot),\psi(0)) \overset{T_{Y}}{\mapsto} (y_{1}(\cdot), y_{1}(-\tau_{0}), \ldots, y_{1}(-\tau_{J+1})) \overset{T_{C}}{\mapsto} (M^{2}_{1}, M^{2}_{0}).
	\end{equation}
	Here $(y_{1}(\cdot), y_{1}(-\tau_{0}), \ldots, y_{1}(-\tau_{J+1}))$ belongs to the Hilbert space $L_{2}([-\tau,0]; \mu_{J}; (\mathbb{C}^{n})^{\otimes 2})$, where $\mu_{J}$ is the sum of the Lebesgue measure and the $\delta$-measures at $-\tau_{j}$ with $j \in \{0,\ldots,J+1\}$. Next, note that $T_{Y}$ and its adjoint $T^{*}_{Y}$ are smoothing in the first component. Specifically, if $\psi(\cdot)$ (resp. $y_{1}(\cdot)$) belongs to $L_{2}$ or $C^{k}$ with $k=0,1,\ldots$, then $y_{1}(\cdot)$ (resp. $\psi(\cdot)$) belongs to $W^{1,2}$ or $C^{k+1}$. Furthermore, if the first component is piecewise $C^{k}$-smooth with discontinuities up to $k$th order at a finite set of points $\mathcal{D}$, then the first component of the image is piecewise $C^{k+1}$-smooth with possible discontinuities at $\mathcal{D} \cup ( -\tau - \mathcal{D} )$. For $T_{Y}$, this is evident right from the differential equation \eqref{EQ: ExplicitTwofoldResolventEquationSystemY1Y2DiffEq}. For $T^{*}_{Y}$, this follows from similar properties of the operator $v \mapsto \int_{-\tau}^{0}Y^{*}_{11}(\theta,s) v(\theta)d\theta$, where $Y_{11}$ is defined in \eqref{EQ: Y1TwofoldResolventIntegralFormula}. We leave details to the interested reader.
	
	From \eqref{EQ: TwofoldDelayIntegralMeasurementClosure}, it can be seen that the action of $T_{C}$ on the first components is as follows. First, it preserves continuity. Second, if the component is $C^{1}$-smooth, then its image is continuous and piecewise $C^{1}$-smooth  with jumps of the derivative at inner delays $-\tau_{j}$. As for $T^{*}_{C}$, it transform an initially continuous function into a discontinuous one, with jumps at the points $-\tau + \tau_{j}$ and $-\tau_{j}$. For higher-order piecewise-smooth functions, it is also evident that the discontinuities are shifted accordingly due to inner delays. However, we shall not provide the exact formulas. In the absence of inner delays, both $T_{C}$ and $T^{*}_{C}$ preserve the class of smoothness.
	
	Since neither $T_{B}$ nor $T^{*}_{B}$ affect the smoothness, the above observations allow us to draw conclusions regarding the smoothing properties of the operators $T_{K} = T_{C}T_{Y}T_{B}$ and $T^{*}_{K} = T^{*}_{B}T^{*}_{Y}T^{*}_{C}$. Specifically, in terms of the first components, both operators take an $L_{2}$-function to a $W^{1,2}$-function, and a continuous function to a continuous piecewise $C^{1}$-smooth function with derivative jumps at the points $-\tau + \tau_{j}$ and $-\tau_{j}$. For smoothing properties of higher order, we leave details to the interested reader.
	
	If $f$ and $\lambda$ are as in the statement, then for any $l \in \{1, \ldots, d_{\mathbb{U}}\}$ the following holds:
	\begin{equation}
		\label{EQ: SmoothnessEigenfunctionDelayCompound}
		\Pi^{(l)}_{\mathbb{U}^{\wedge}_{2}}\left(\Sigma^{(\kappa)}(p) + \sum_{i=1}^{d_{\mathbb{U}}}\kappa_{i}\Pi^{(i)}_{\mathbb{U}^{\wedge}_{2}}\right)f = (\lambda + \kappa_{l}) \Pi^{(l)}_{\mathbb{U}^{\wedge}_{2}}f.
	\end{equation}
	According to \eqref{EQ: DelayCompoundSiLipschitz}, \eqref{EQ: DelayCompoundSiMonotone}, and \eqref{EQ: DelayCompoundSigmaKappaSum}, the action of the operator from the left-hand side in \eqref{EQ: SmoothnessEigenfunctionDelayCompound} consists in moving $f$ through the actions of $T_K$ and $T_K^*$, as well as their compositions with operators that act on the level of function values and thus do not affect smoothness. Since $\lambda + \kappa_{l} \not= 0$ and $f_{1}$ is initially an element of $L_{2}$, it follows from the above that the first component of $\Pi^{(l)}_{\mathbb{U}^{\wedge}_{2}}f$ is a continuous piecewise $C^{1}$-smooth function with possible derivative jumps at the points $-\tau + \tau_{j}$ and $-\tau_{j}$ corresponding to inner delays $\tau_{j}$. Since $f$ is the sum over $l$ of such projections, this establishes the desired statement.
\end{proof}

\subsection{Truncations of $T_{K}$ and numerical schemes}
\label{SUBSEC: TruncationsOfTransferOperators}
Consider a family functions $\phi_{k}(\theta)$, where $k \in \mathbb{Z}$, which form an orthonormal basis in $L_{2}(-\tau,0;\mathbb{C})$. Specific choices will be discusses below. From this, we can construct orthonormal bases in $\mathbb{H}_{\mathbb{U}}$ and $\mathbb{H}_{\mathbb{M}}$ as follows. Let $e_{1},\ldots, e_{n}$ be the standard basis in $\mathbb{C}^{n}$, and let $u_{1},\ldots, u_{d_{\mathbb{U}}}$ and $m_{1},\ldots,m_{d_{\mathbb{M}}}$ be orthonormal bases in $\mathbb{U}$ and $\mathbb{M}$, respectively. Then the elements of $\mathbb{H}_{\mathbb{U}}$:
\begin{equation}
	\label{EQ: TrigonometricBasis}
	\{(\phi_{k}(\cdot) \otimes (e_{i} \otimes u_{j}), 0)\} \quad \text{and} \quad (0, (e_{i} \otimes u_{j}))
\end{equation}
taken over all $k \in \mathbb{Z}$, $i \in \{1,\ldots,n\}$, and $j \in \{1,\ldots, d_{\mathbb{U}}\}$, form an orthonormal basis. By exchanging $u_{j}$ with $m_{j}$ in the above construction, we obtain an orthonormal basis for $\mathbb{H}_{\mathbb{M}}$.

For computations, we fix some integer $N \geq 0$ and restrict the index $k$ in the basis elements to a finite set $\mathcal{I}_{N}$, which monotonically increases in $N$ and exhausts $\mathbb{Z}$ as $N \to \infty$. Let $P^{\mathbb{U}}_{N}$ and $P^{\mathbb{M}}_{N}$ denote the orthogonal projectors onto the corresponding subspaces in $\mathbb{H}_{\mathbb{U}}$ and $\mathbb{H}_{\mathbb{M}}$, respectively. This provides finite-dimensional truncations $W_{N} \coloneq P^{\mathbb{M}}_{N}T_{K}P^{\mathbb{U}}_{N}$ of $T_{K}$, which can be associated with a ($n d_{\mathbb{M}} |\mathcal{I}_{N}|\times n d_{\mathbb{U}} |\mathcal{I}_{N}|$)-matrix.

Exchanging $T_{K}$ and $T^{*}_{K}$ in \eqref{EQ: DelayCompoundSiLipschitz} and \eqref{EQ: DelayCompoundSiMonotone} with $W_{N}$ and $W^{*}_{N}$, respectively, we obtain truncations $\Sigma^{(\kappa)}_{N}$ of the operator $\Sigma^{(\kappa)}$ from \eqref{EQ: DelayCompoundSigmaKappaSum}.

In the context \eqref{EQ: FrequencyInequalityCompoundDelayMixedConstraint}, let $\alpha_{N}(\omega)$ be the largest eigenvalue of $\Sigma^{(\kappa)}_{N}(p)$ with $p=-\nu_{0} + i \omega$. From the general perspective, $\alpha_{N}(\omega)$ monotonically converges to $\alpha(\omega)$ as $N \to \infty$ uniformly in $\omega$ from compact intervals. Moreover, the rate of convergence depends on the decay of Fourier coefficients of the principal eigenfunction $f$.
\begin{remark}
	More rigorously, in our case the rate of convergence also depends on the decay of Fourier coefficients of the images $T_{K}f$ and $T^{*}_{K}T_{K}f$, since the truncations of $\Sigma^{(\kappa)}_{N}$ involve the multiplicative terms $P^{\mathbb{U}}_{N} T^{*}_{K} P^{\mathbb{M}}_{N} T_{K} P^{\mathbb{U}}_{N}$ with the truncation taken in the middle. However, according to the proof of Proposition \ref{PROP: PrincipalEigenfunctionDelayCompound}, the images has the same regularity properties as $f$.
\end{remark}

For comparison, we first take the standard trigonometric basis $\phi_{k}(\theta) = e^{i2\pi\theta k /\tau}/\sqrt{\tau}$, where $k \in \mathbb{Z}$. Since the principal eigenfunction of $\Sigma^{(\kappa)}$ does not in general have equal values at the endpoints $-\tau$ and $0$, the $k$th Fourier coefficient decays only as $O(1/|k|)$. Overall, this gives the convergence of $\alpha_{N}(\omega)$ to $\alpha(\omega)$ with the order of $O(N^{-1/2})$, where $\mathcal{I}_{N} = \{-N,\ldots,N\}$. To increase the rate of convergence, we complement the trigonometric basis in the form of sines and cosines by a linear function, and then orthonormalize the resulting family. This determines the enriched trigonometric basis. We omit the cumbersome formulas and a proof of the fact that the rate of convergence in this basis is $O(N^{-3/2})$, where $\mathcal{I}_{N}$ contains the indices for the constant and linear functions, as well as trigonometric functions with frequency factors up to $\pm N$. We only note that the presence of inner delays, causing discontinuities of the first derivative in accordance with Proposition \ref{PROP: PrincipalEigenfunctionDelayCompound}, slows down the convergence, but preserves the asymptotics.

For computations, we discretize the interval $[-\tau,0]$ by the uniform grid of $N_{\operatorname{grid}}$ points. For convenience, all inner delays are adjusted to the nearest grid points. The kernels in \eqref{EQ: TransferOperatorTwoFoldDelayIntegralRepresentation} are evaluated on the grid. Numerical integration when calculating truncations is performed according to the trapezoidal rule.

Let us briefly discuss the prospects for rigorous numerical methods that combine interval arithmetic with explicit error estimates. First of all, we have, in theory, access to the regularity properties of the kernels, both with respect to the spatial variables $(\theta, s)$ and with respect to the spectral variable $p$, and we can estimate the error in their computation from the matrix exponential and explicit formulas. In particular, we can estimate the error involved in calculating truncations. In turn, the error in approximation by $\alpha_{N}$ can be assessed by deriving explicit formulas for the decay of Fourier coefficients in terms of the kernel components, at least in some cases. Another approach for this is to estimate the accuracy of the full truncation using the optimization of Schur test functions \cite{AnikushinRomanov2025Schur}. However, for the operators of Hardy--Volterra type, this results in the worst asymptotics $O(N^{-1/2})$, determined by the convergence of truncations in the Hilbert--Schmidt norm. Finally, for the computation of eigenvalues of Hermitian matrices, there exist rigorous numerics (see, e.g., \cite{ImakuraMorikuniTakayasu2023}).

Summarizing the above, it is theoretically possible to rigorously test frequency inequalities over a \textit{finite} interval of $\omega$. For the infinite interval, the problem arises of investigating the behavior of $\alpha(\omega)$ as $|\omega| \to \infty$. In \cite{Anikushin2023Comp, AnikushinRomanov2023FreqConds}, it is conjectured that $\alpha(\omega)$ behaves asymptotically like a uniformly almost periodic function or even tends toward a constant (the latter being more likely in our case). For scalar equations with a single Lipschitz block, this was proved in \cite{AnikushinRomanov2023FreqConds} using explicit formulas for the kernels. Here, the asymptotic constant is expressed in terms of the norm of an integral operator with an asymptotic kernel, and the convergence is of order $O(1/|\omega|)$, where the constant can be estimated explicitly. In the general case, there should be more universal ways to access such properties.

In any case, numerical experiments show that the most interesting behavior is indeed observed only within a small interval of $\omega$ (which aligns with engineering intuition), and the eigenvalues $\alpha_{N}(\omega)$, computed using the enriched basis, converge very rapidly and uniformly over this interval: $N=5$ is sufficient to achieve accuracy on the order of thousandths.

\subsection{Package utilization}
\label{SUBSEC: PackageUtilization}
In the repository, we provide a package with the main class \texttt{QuadraticBendixsonDDE}, which implements numerical methods for verifying frequency inequalities. Its use is demonstrated by the examples considered in Section \ref{SEC: Examples}. In this section, we provide an overview of its main capabilities.

First, the user must create an instance of the class:
\begin{lstlisting}[language=Python]
wrapper = QuadraticBendixsonDDE(A0, Atau, B, Cs, taus, 
	constr, kappas, tau, Q_U, Q_M)
\end{lstlisting}
where the arguments are described as follows:
\begin{description}
	\item[\texttt{A0}, \texttt{Atau}, and \texttt{B}]: The matrices $A_{0}$, $A_{\tau}$, and $\widetilde{B}$, respectively, which determine the linear block in \eqref{EQ: DelayNonautonomous};
	\item[\texttt{Cs} and \texttt{taus}]: Lists of the matrices $C_{\tau_{j}}$ and the delays $\tau_{j}$ (with any sign), respectively, which determine the measurement operator $C$ in \eqref{EQ: MeasurementOperator}. Can be passed in any order within the list, but the indices of $C_{\tau_{j}}$ and $\tau_{j}$ must match;
	\item[\texttt{constr}]: A list of constraints, \nameref{DESC: LipC} or \nameref{DESC: MonC}, for each feedback component $F'_{i}$ from \eqref{EQ: DelayFeedbackComponentsDistinguished}. Each constraint is a dictionary containing the following keys:
	\begin{description}
		\item[\texttt{'type'}]: The type of constraint, \texttt{'Lip'} for Lipschitz and \texttt{'Mon'} for monotone;
		\item[\texttt{'Lam'}]: The constraint constants. Must be a single value $\Lambda^{(i)}$ for Lipschitz constraints and a tuple $(\Lambda^{(i)}_{1},\Lambda^{(i)}_{2})$ for monotone constraints;
		\item[\texttt{'Proj\_U'}]: A matrix determining the orthogonal projector $\Pi^{(i)}_{\mathbb{U}}$;
		\item[\texttt{'Proj\_M'}]: A matrix determining the orthogonal projector $\Pi^{(i)}_{\mathbb{M}}$;
		\item[\texttt{'h'}]: A matrix determining the mapping $h_{i}$ (for monotone constraints only);
		\item[\texttt{'h\_inv'}]: A matrix determining the mapping $h^{(-1)}_{i}$ (for monotone constraints only);
	\end{description}
	If there is only one constraint, it suffices to add just the \texttt{'type'} and \texttt{'Lam'} entries.
	\item[\texttt{kappas}]: The list of initial values for the coefficients $\kappa_{i}$ from \eqref{EQ: DelayCompoundSigmaKappaSum}. The default value is \texttt{None}, meaning that the coefficients will be initialized as $\kappa_{i} = 1/n_{\mathbb{U}}$ for any $i$;
	\item[\texttt{tau}]: The value of $\tau$ determining the interval $[-\tau,0]$. The default value is $1$;
	\item[\texttt{Q\_U} and \texttt{Q\_M}]: The symmetric positive-definite real matrices determining the inner products in $\mathbb{U}$ and $\mathbb{M}$, respectively. The default values are \texttt{None}, meaning that the matrices are initialized as identities.
\end{description}

To initiate the computations, the user must call the following procedure:
\begin{lstlisting}[language=Python]
vals = wrapper.computeCriterion(
    ps, Ns_trunc, basis, N_grid, findBestKappas, verbose, 
    n_jobs, eig_solver, device, block_size, block_mem_mb)
\end{lstlisting}
where the arguments are described as follows:
\begin{description}
	\item[\texttt{ps}]: A list of complex values $p$ for which approximations are calculated;
	\item[\texttt{Ns\_trunc}]: A list of $N$ for which truncations are calculated;
	\item[\texttt{basis}]: The type of basis to use in computations, \texttt{'Trigonometric'} for the trigonometric basis and \texttt{'Enriched'} (default) for the enriched trigonometric basis;
	\item[\texttt{N\_grid}]: A number of points in the uniform grid discretizing the interval $[-\tau,0]$. Note that each inner delay $-\tau_{j}$ will be adjusted to the nearest grid point. The default value is $251$;
	\item[\texttt{findBestKappas}]: If set to \texttt{True}, the maximum of $\alpha_{N}(p)$ over $p$ in \texttt{ps} will be minimized\footnote{This is done via the SLSQP method from SciPy.} over $\kappa$. For the optimization, $N$ is taken as the minimum from the list \texttt{Ns\_trunc}. The optimized $\kappa$ is memorized in the class instance and is used to compute $\alpha_{N}(p)$ for all $N$ in \texttt{Ns\_trunc} in the present call. This requires the storage of the truncation matrices over \texttt{ps}, which occupies about $128 L (2N_{\operatorname{max}}+1)^{2} n^{2} d_{\mathbb{M}} d_{\mathbb{U}}$ bits of memory, where $L$ is the length of \texttt{ps} and $N_{\operatorname{max}}$ is the maximum over \texttt{Ns\_trunc}. The default value is \texttt{False};
	\item[\texttt{verbose}]: If set to \texttt{True}, displays various messages regarding intermediate calculations. The default value is \texttt{False};
	\item[\texttt{n\_jobs}]: The number of CPU cores (workers) to use in parallel computations. When \texttt{findBestKappas} is set to \texttt{True}, the memory usage is multiplied by this parameter. Works only with \texttt{device='cpu'}. The default value is $1$;
	\item[\texttt{eig\_solver}]: The eigenvalue solver to use. Can be \texttt{'lanczos'} (default) or \texttt{'full'}. This parameter is ignored if \texttt{device='gpu'};
	\item[\texttt{device}]: The device used for the main computations. Can be \texttt{'cpu'} (default) or \texttt{'gpu'}; 
	\item[\texttt{block\_size}]: Works only when \texttt{findBestKappas} is set to \texttt{False}. Determines the number of points in \texttt{ps} taken in a block for computing truncation matrices and their eigenvalues. So, the overall memory usage is proportional to the size of a block, not the length of \texttt{ps}. The default value is \texttt{None}, meaning that the size is determined from \texttt{block\_mem\_mb};
	\item[\texttt{block\_mem\_mb}]: When \texttt{block\_size} is set to \texttt{None}, determines the block size in MBs. The default value is \texttt{None}, meaning that the standard values are used: 256 MB per worker on CPU and 1024 MB on GPU;
	\item[\texttt{returns}]: A list of the same length as \texttt{Ns\_trunc} whose $i$th element is a list of approximating largest eigenvalues $\alpha_{N_{i}}(p)$ for $p$ in \texttt{ps}, where $N_{i}$ = \texttt{Ns\_trunc[i]}.
\end{description}

For searching the optimal delay scaling in the case of systems of equations (see Remark \ref{REM: DelayScalingNoninvariance}), the class offers the following method:
\begin{lstlisting}[language=Python]
vals, opt_scal = wrapper.findBestScaling( ps, N_trunc, 
    basis, N_grid, findBestKappas, verbose, n_jobs, 
    eig_solver, device, block_size, block_mem_mb)
\end{lstlisting}
where \texttt{N\_trunc} is a single number $N$ used to compute truncations. The other parameters are as in the previous method and are used for its internal calls, except for \texttt{verbose} which affects only the outer messages. The method returns the array of values $\alpha_{N}(p)$ for $p$ in \texttt{ps} and the optimal scaling, which automatically applies to the class parameters. When \texttt{findBestKappas} is set to \texttt{True}, then the optimal $\kappa$ is also applied to the class parameters.

Since the dependence on the scaling parameter and $\kappa$ is not smooth and possesses multiple local minima, the procedure uses the bounded Brent's method, implemented in SciPy. For efficient computations, it is recommended to use about $40$ points in \texttt{ps}, $N=5$, and the default (or smaller) \texttt{N\_grid}. In this case, the calculations are better suited for CPU.

The core code logic (approximately 1000 lines) was initially written and tested by the authors. We then used the Claude Sonnet 5.5 and Grok 4.5 AI models to optimize the code for CPU and GPU computations, respectively.

%% file: Hopfield.tex
\section{Examples}
\label{SEC: Examples}
In this section, we consider two examples studied in our works \cite{AnikushinRomanov2023FreqConds,AnikushinRomanovQBC2026}, where explicit formulas for the kernels of integral operators are used. We show that the results of the general computational scheme presented in the package are consistent with the results presented in these papers. Furthermore, we refine the corresponding estimates for the boundary of global stability by leveraging the faster convergence achieved through the enriched basis.

\subsection{Delayed Hopfield networks}
\label{SUBSEC: HopfieldConcrete}
We consider the following Hopfield network with specific parameters:
\begin{equation}
	\label{EQ: HopfieldSpecificParameters}
	\begin{split}
		\dot{x}_{1}(t) &= -1.1x_{1}(t) - 0.2\sigma_{1}(x_{1}(t-\tau_{1})) + \sigma_{2}(x_{2}(t-\tau_{2})),\\
		\dot{x}_{2}(t) &= -1.2x_{2}(t) + \sigma_{1}(x_{1}(t-\tau_{1})) -0.1 \sigma_{2}(x_{2}(t-\tau_{2})),
	\end{split}
\end{equation}
where the sigmoid nonlinearities $\sigma_{1}$ and $\sigma_{2}$ are given by
\begin{equation}
	\label{EQ: HopfieldLastExample}
	\sigma_{1}(y) = \frac{2}{\pi}\arctan\left(\frac{1.4\pi}{2}y\right) \quad \text{and} \quad \sigma_{2}(y) = \frac{2}{\pi}\arctan\left( \frac{1.5\pi}{2}y\right).
\end{equation}

It can be shown that \eqref{EQ: HopfieldSpecificParameters} generates a dissipative semiflow $\pi$ in $C([-\tau,0];\mathbb{R}^{2})$ with a global attractor $\mathcal{A}$. From the perspective of global stability problems, the system \eqref{EQ: HopfieldSpecificParameters} has 3 equilibria: $\phi^{0} \equiv (0,0)^{T}$ and $\phi^{\pm} \equiv (0.2139...,0.2090...)^{T}$. In particular, none of the methods designed for cases with a single equilibrium can be applied here. 

We have taken this example from \cite[Section 5]{YeMichelWang1994}, where the boundary of global stability for the case of identical delays $\tau_{1}=\tau_{2}=\tau$ was estimated as $\tau < 0.579$. This estimate was improved in our work \cite{AnikushinRomanovQBC2026} to $\tau \leq 1.2$ using the quadratic Bendixson criterion with truncations taken in the standard trigonometric basis. For simplicity, \cite{AnikushinRomanovQBC2026} considers only single vector blocks in the feedback part, which precludes the use of distinct bounds for the derivatives. In our case, we are able to incorporate different constants into the frequency inequalities, thereby refining the estimate to $\tau \leq 1.4$, see below.

It should be noted that for $\tau > \tau^{*}_{0} \approx 1.93$, the primary saddle equilibrium $\phi^{0}$ acquires a focus-type instability, whereas $\phi^{\pm}$ remain linearly stable until $\tau = \tau^{*}_{\pm} \approx 3.61$. In particular, the bound in $\tau$ for the uniform area contraction, determined by the quadratic Bendixson criterion, cannot exceed $\tau^{*}_{0}$. Moreover, \cite{AnikushinRomanovQBC2026} reports the presence of hidden attracting periodic orbits in \eqref{EQ: HopfieldSpecificParameters} for $\tau \in [2.11, 2.16]$. Thus, it is possible that the boundary of global stability is determined by nonlocal bifurcations or, following the terminology of \cite{Kuzetal2020Lorenz}, is hidden.

The linearization of \eqref{EQ: HopfieldSpecificParameters} over $(\mathcal{A},\pi)$ fits into the framework of \eqref{EQ: DelayNonautonomous} as follows:
\begin{equation}
	\label{EQ: HopfieldExampleABCF}
	\begin{split}
		A_{0} = \begin{pmatrix}
			-1.1 & 0\\
			0 & -1.2
		\end{pmatrix}, \ A_{\tau} = \begin{pmatrix}
		0 & 0\\
		0 & 0
		\end{pmatrix},  \ \widetilde{B} = \begin{pmatrix}
			-0.2 & 1\\
			1 & -0.1
		\end{pmatrix}, \ \mathbb{U} = \mathbb{M} = \mathbb{R}^{2},\\
		C_{\tau_{1}} = \begin{pmatrix}
			1 & 0\\
		\end{pmatrix}, \ C_{\tau_{2}} = \begin{pmatrix}
			0 & 1\\
		\end{pmatrix}, \
		F'(\wp) = \begin{pmatrix}
			\sigma'_{1}(\wp_{1}(-\tau_{1})) & 0\\
			0 & \sigma'_{2}(\wp_{2}(-\tau_{2}))
		\end{pmatrix},
	\end{split}
\end{equation}
where $\wp=(\wp_{1},\wp_{2}) \in C([-\tau,0];\mathbb{R}^{2})$. Moreover, $F'(\wp)$ can represented in the block form \eqref{EQ: DelayFeedbackComponentsDistinguished} with $n_{\mathbb{U}}=2$, $F'_{1}(\wp) = \sigma'_{1}(\wp_{1}(-\tau_{1}))$, $F'_{2}(\wp) = \sigma'_{2}(\wp_{2}(-\tau_{2}))$,
\begin{equation}
	\label{EQ: HopfieldExampleBlock}
	\begin{split}
		\mathbb{M}^{1}=\mathbb{U}^{1} = \mathbb{R} \times \{0\}, \ \mathbb{M}^{2}=\mathbb{U}^{2} =  \{0\} \times \mathbb{R},\\
		\Pi^{(1)}_{\mathbb{U}} = \Pi^{(1)}_{\mathbb{M}} = \begin{pmatrix}
			1 & 0\\
			0 & 0
		\end{pmatrix}, \ \Pi^{(2)}_{\mathbb{U}} = \Pi^{(2)}_{\mathbb{M}} = \begin{pmatrix}
		0 & 0\\
		0 & 1
		\end{pmatrix}.
	\end{split}
\end{equation}

The blocks $F'_{1}$ and $F'_{2}$ satisfy \nameref{DESC: MonC} with constants $\Lambda^{(1)}_{1} = \Lambda^{(2)}_{1} = 0$, $\Lambda^{(1)}_{2} = 1.4$, $\Lambda^{(2)}_{2} = 1.5$, and
\begin{equation}
	\label{EQ: HopfieldExampleMonotoneParams}
	h_{1}=h^{(-1)}_{1} = h_{2}=h^{(-1)}_{2} = \begin{pmatrix}
		1 & 0\\
		0 & 1
	\end{pmatrix}.
\end{equation}

\begin{figure}[t]
	\begin{minipage}{.5\textwidth}
		\includegraphics[width=\textwidth,angle=0]{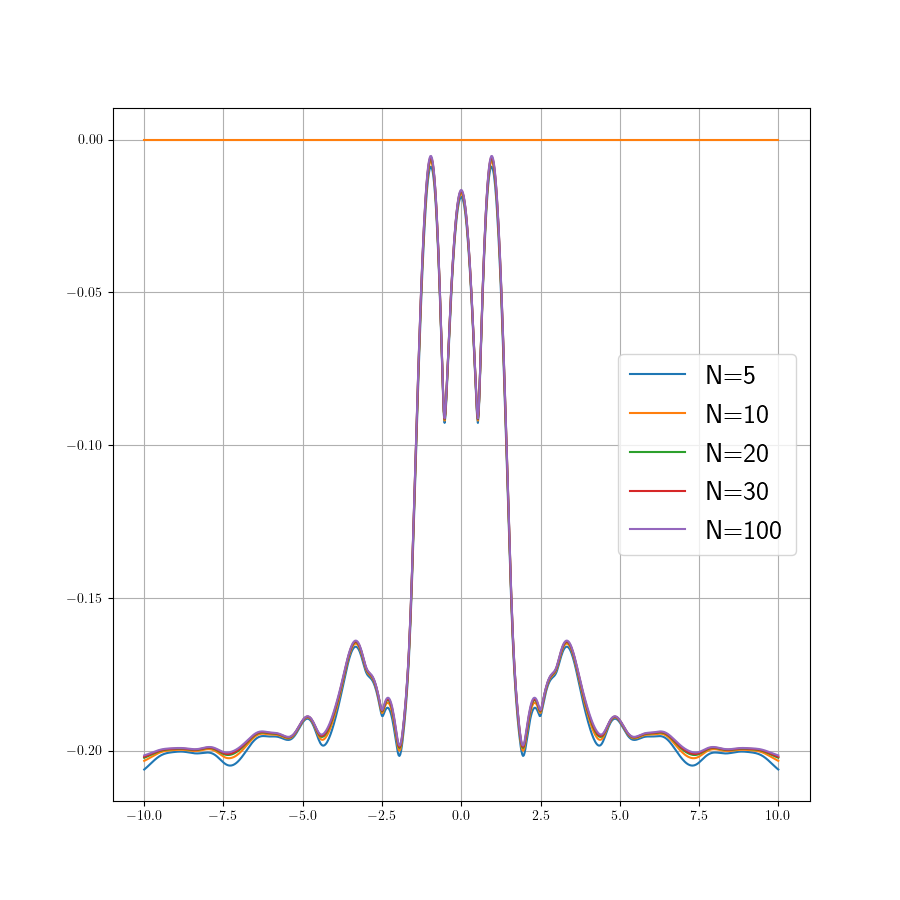}
	\end{minipage}%
	\begin{minipage}{.5\textwidth}
		\includegraphics[width=\textwidth,angle=0]{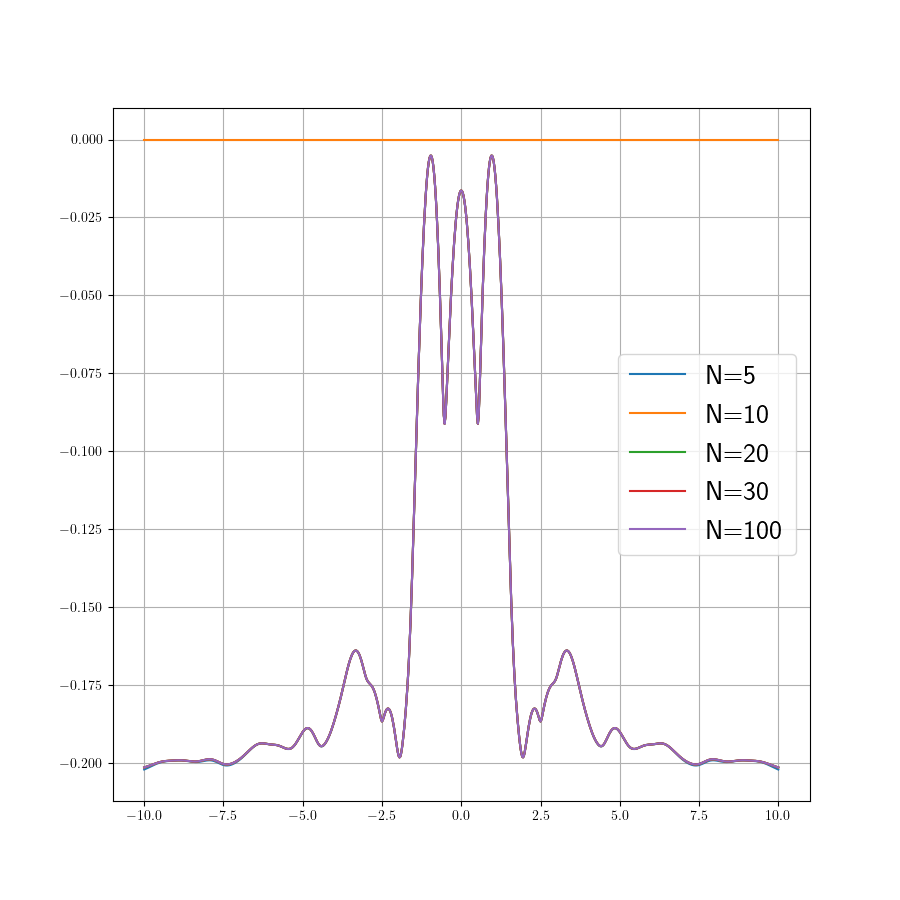}
	\end{minipage}%
	\caption{Plots of $\alpha_{N}(\omega)$ versus $\omega$ for the Hopfield model \eqref{EQ: HopfieldSpecificParameters} with $\tau_{1}=\tau_{2} = 1.4$ whose linearization is considered as in \eqref{EQ: HopfieldExampleABCF}, \eqref{EQ: HopfieldExampleBlock}, and \eqref{EQ: HopfieldExampleMonotoneParams}, and then applying the scaling parameter $\varkappa=0.675$. Truncations are taken with respect to the trigonometric basis (left) and the enriched trigonometric basis (right) with $N=5$ (blue), $N=10$ (orange), $N=20$ (green), $N=30$ (red), and $N=100$ (purple). See the repository for implementation details.}
	\label{FIG: HopfieldQBCTwoInputTruncations}
\end{figure}

Clearly, $A_{0}$ and $A_{\tau}$ determine a stable linear system. Thus, Theorem \ref{TH: ExpontialStabilityTwofoldDelayGeneral} can be applied with $\nu_{0} = 0$. As in \cite{AnikushinRomanovQBC2026}, we also use a nontrivial delay scaling $t \mapsto \varkappa t$, which for $\tau=1.4$ reduces to the choice of $\varkappa=0.675$. Figure \ref{FIG: HopfieldQBCTwoInputTruncations} shows the corresponding graphs of $\alpha_{N}(\omega)$ obtained by the general subroutine in this case. It can be seen that, within the enriched basis, choosing $N=5$ is already sufficient to obtain approximations that are indistinguishable from those obtained with $N=100$. The figures are consistent with \cite[Figure 1]{AnikushinRomanovQBC2026}. In computations, we discretized the interval $[-\tau,0]$ by taking $N_{\operatorname{grid}}=1000$ points in the uniform grid. Increasing the number of points to $5000$ did not change the gap $\approx 5 \cdot 10^{-3}$ between the graph of $\alpha_{N}(\omega)$ and the threshold line. In fact, our method indicates the global stability up to $\tau = 1.45$, where the gap becomes $\approx 8 \cdot 10^{-4}$.

The interval $[0,10]$ of $\omega$ was discretized by the uniform grid of $401$ points. On our device\footnote{It is CPU AMD Ryzen 5 5600 OEM with 32GB 3200 MHz DDR4 RAM and GPU Palit GeForce RTX 4060 Dual.}, GPU computations took 54 seconds for the full set of $N$ and 6.5 seconds only for $N=5$. For comparison, CPU computations with \texttt{n\_jobs=4} took 100 seconds and 4.5 seconds, respectively.

\subsection{Mackey--Glass equations}

In this section, we demonstrate applications of the quadratic Bendixson criterion by means of the Mackey--Glass model:
\begin{equation}
	\label{EQ: MackeyGlassExample}
	\dot{x}(t) =  - \gamma x(t) + \beta \frac{ x(t-\tau) }{ 1 + |x(t-\tau)|^{\kappa}  },
\end{equation}
where $\tau, \beta,\gamma > 0$, and $\kappa > 1$ are real parameters. It can be shown that the model generates a dissipative semiflow $\pi$ in $C([-\tau,0];\mathbb{R})$ with a global attractor $\mathcal{A}$, see \cite[Lemma 6.3]{Anikushin2023LyapExp}. According to the physiological perspective, one is interested in the dynamics of \eqref{EQ: MackeyGlassExample} restricted to the invariant cone of positive functions, where the system admits a unique equilibrium $\phi^{+} \equiv (\beta\gamma^{-1}-1)^{1/\kappa}$ for $\beta > \gamma$.

Since frequency inequalities for scalar equations are invariant with respect to delay scaling, it is convenient to normalize the delay in \eqref{EQ: MackeyGlassExample} by applying the scaling $t \mapsto \tau t$. After the scaling, it can be verified that the linearization of \eqref{EQ: HopfieldSpecificParameters} over $(\mathcal{A},\pi)$ fits into the framework of \eqref{EQ: DelayNonautonomous} as follows:
\begin{equation}
	\label{EQ: MackeyGlassLinearizationFit}
	\begin{split}
		A_{0} = -\tau \gamma, \ A_{\tau} = -\tau\beta(\Lambda-1), \ \widetilde{B} = \tau \beta, \mathbb{U}=\mathbb{M}=\mathbb{R},
		C_{\tau} = 1, \\
		F'(\wp) = f'(\wp(-1)) + \Lambda  - 1,
	\end{split}
\end{equation}
where $f(y) = y/(1+|y|^{\kappa})$, and $\Lambda$ is given by
\begin{equation}
	\label{EQ: MackeyGlassStabilityLambdaDef}
	\Lambda = \frac{1}{2} \left( \frac{(\kappa-1)^{2}}{4\kappa} + 1 \right).
\end{equation}
It can be checked that $|F'(\wp)| \leq \Lambda$, that is, $F'$ determines a scalar feedback satisfying the Lipschitz constraint \nameref{DESC: LipC} with the constant $\Lambda$.

Next, we consider the classical parameters $\gamma = 0.1$, $\beta = 0.2$, and $\kappa = 10$. In this case, $\phi^{+}$ loss the local stability for $\tau$ exceeding the parameter\footnote{In \cite[Equation (5.9)]{AnikushinRomanov2023FreqConds}, the parameter $\tau^{*}$ is incorrectly determined as $\tau^{*} = (8/3)\arccos(-1/4) \approx 4.86$.}
\begin{equation}
	\label{EQ: MackeyGlassBifurcationParameter}
	\tau^{*} = \frac{10}{\sqrt{15}}\arccos\left( -\frac{1}{4} \right) = 4.708\ldots,
\end{equation}
corresponding to the Andronov--Hopf bifucation. It is expected that $\phi^{+}$ is globally stable in the cone of positive functions for all $\tau < \tau^{*}$.

\begin{remark}
	It is well known that $\dot{x}(t)= -ax(t) - b x(t-1)$ with $a,b > 0$ is stable if and only if $b < \sqrt{a^{2} + \xi^{2}}$, where $\xi \in (\pi/2,\pi)$ is the unique root of $\xi \cot \xi = - a$. Then for $a = -A_{0}$ and $b=-A_{\tau}$ as in \eqref{EQ: MackeyGlassLinearizationFit} and the classical choice of parameters, we obtain stability for all $\tau < \frac{\pi - \operatorname{arctg}0.225}{0.0225} \approx 129.8$.
\end{remark}

According to the above remark, $A_{0}$ and $A_{\tau}$ determine a stable linear system for $\tau < \tau^{*}$. In particular, Theorem \ref{TH: ExpontialStabilityTwofoldDelayGeneral} can be applied with $\nu_{0} = 0$. Figure \ref{FIG: HopfieldQBCTwoInputTruncations} shows the corresponding graphs of $\alpha_{N}(\omega)$ obtained by the general subroutine for $\tau = 4.63$. Here, the gap between the graphs and the threshold line is $3 \cdot 10^{-3}$, and one can draw conclusions similar to those given in the previous example. Moreover, the figures are consistent with \cite[Figures 2 and 4]{AnikushinRomanov2023FreqConds}. Overall, within the enriched basis, we refine the estimate $\tau \leq 4.5$ (resp. 4.55) reported in \cite{AnikushinRomanov2023FreqConds} (resp. \cite{AnikushinRomanov2025Schur}). It should be noted that $\tau=4.63$ is very close to the bifurcation parameter $\tau^{*}$ from \eqref{EQ: MackeyGlassBifurcationParameter}.

We used $401$ points to discretize the interval $[0,35]$ of $\omega$ and $N_{\operatorname{grid}}=1000$ points in the uniform grid on $[-1,0]$. In this case, GPU calculations took 9 seconds for the full set of $N$ and 5 seconds only for $N=5$. For comparison, CPU calculations (with \texttt{n\_jobs=4}) took 9 and 2.5 seconds, respectively.

As in \cite{AnikushinRomanov2023FreqConds}, the result can be compared with the generalized Myshkis criterion of \cite{Lizetal2003}, which in our case gives the condition $\tau < -10 \left[ \ln 4 + \ln\ln(20/17) \right] = 4.306\ldots$. While the method of \cite{Lizetal2003} is specific to scalar equations with a single equilibrium and unimodal nonlinearities, our approach uses only the derivative bound.

It should also be noted that the Popov criterion \cite{PopovKhalanay1962} is not applicable for the parameters under consideration, regardless of the value of $\tau$. However, one can apply the usual quadratic criterion in a tricky way. For this, one should rewrite \eqref{EQ: MackeyGlassExample} for a given nonstationary solution $x(\cdot)$ in the positive cone as follows:
\begin{equation}
	\dot{y}(t) = -\gamma y(t) - g(t) y(t-\tau),
\end{equation}
where $y(t) = x(t) - x^{+}$, $x^{+} = (\beta\gamma^{-1}-1)^{1/\kappa}$, and $g(t) = -\beta (f(x(t-\tau))-f(x^{+}))/y(t-\tau)$. As in \cite{LizRost2009}, one can utilize the unimodality to iteratively obtain bounds for the attractor of \eqref{EQ: MackeyGlassExample} in the positive cone. For $g$, this gives sector bounds: $g(t) \in [\Lambda_{1},\Lambda_{2}]$ for some $0 < \Lambda_{1} \leq \Lambda_{2}$, provided that $x(t)$ lies on the attractor. Then the usual quadratic criterion (see, e.g., \cite[Theorem 4.3]{Anikushin2020FreqDelay}) can be applied to study the absolute stability of
\begin{equation}
	\dot{y}(t) = -\gamma y(t) + \Lambda_{1}y(t-\tau) - G(t) y(t-\tau)
\end{equation}
in the class of $G(t)$ restricted to the sector $[0, \Lambda]$ with $\Lambda = \Lambda_{2}-\Lambda_{1}$. For the classical parameters, this ensures stability for $\tau \leq 4.54$. We omit the details.

\begin{figure}[t]
	\begin{minipage}{.5\textwidth}
		\includegraphics[width=\textwidth,angle=0]{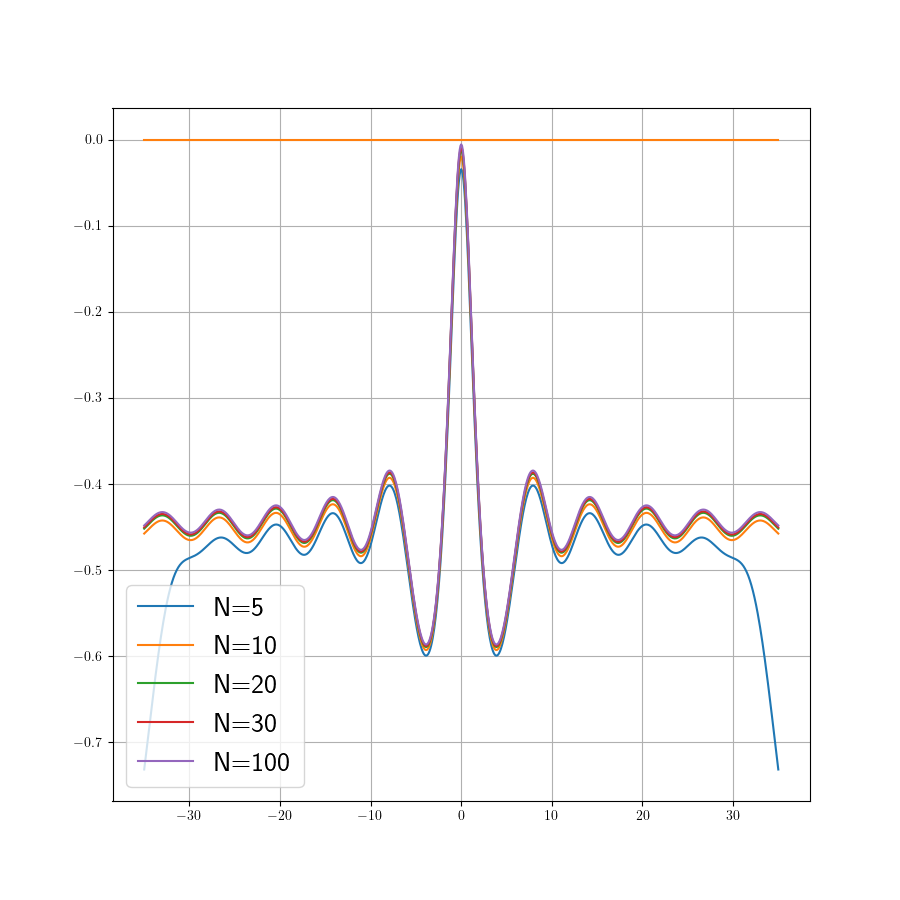}
	\end{minipage}%
	\begin{minipage}{.5\textwidth}
		\includegraphics[width=\textwidth,angle=0]{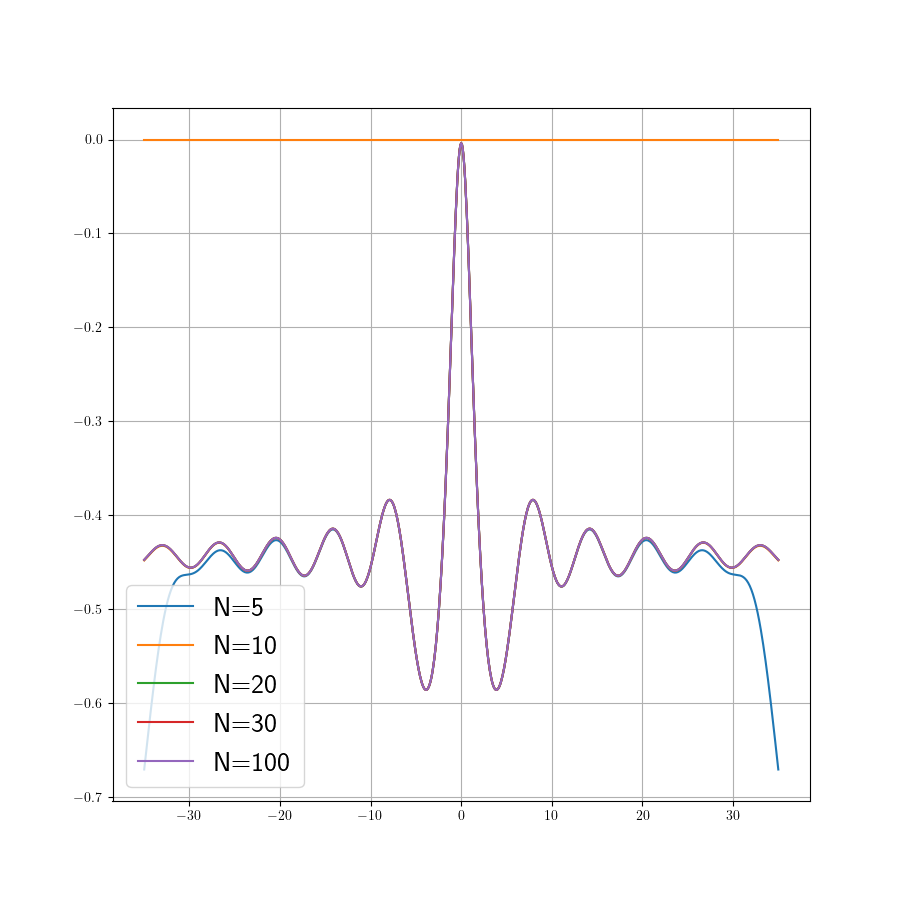}
	\end{minipage}%
	\caption{Plots of $\alpha_{N}(\omega)$ versus $\omega$ for the Mackey--Glass model \eqref{EQ: MackeyGlassExample} with $\gamma = 0.1$, $\beta = 0.2$, $\kappa = 10$, and $\tau = 4.63$ whose linearization is considered as in \eqref{EQ: MackeyGlassLinearizationFit} after the delay normalization. Truncations are taken with respect to the trigonometric basis (left) and the enriched trigonometric basis (right) with $N=5$ (blue), $N=10$ (orange), $N=20$ (green), $N=30$ (red), and $N=100$ (purple). See the repository for implementation details.}
	\label{FIG: MackeyGlassQBCTruncations}
\end{figure}